\documentclass[10pt, aps, pra, nofootinbib, 
               amsmath, amssymb, twocolumn,
               preprintnumbers, showpacs,
               raggedbottom,
               floatfix]{revtex4-1}

\usepackage[T1]{fontenc} 

\usepackage{etoolbox}
\usepackage{etextools}
\usepackage[acronym]{glossaries}
\glsdisablehyper
\makeglossaries

\usepackage[utf8]{inputenc}

\usepackage[breaklinks]{hyperref}
\usepackage{graphicx}
\graphicspath{{./figures/arXiv/}}
\usepackage{dcolumn}
\usepackage{calc}

\usepackage{lettrine}

\usepackage[tracking]{microtype}
\SetTracking{ encoding = *, shape = sc }{ 25 }
\usepackage{xcolor}

\usepackage[sc,osf]{mathpazo}
\usepackage[euler-digits,small]{eulervm}
\AtBeginDocument{\renewcommand{\hbar}{\hslash}}

\usepackage{bm}

\LettrineOptionsFor{T}{
  lines=3,
  loversize=0.07,
  lraise=-0.07,
  lhang=0.46,
  findent=0.4em,
  nindent=-0.0\LettrineWidth-0.5em}

\usepackage{booktabs}

\usepackage{todonotes}
\usepackage{pdfcomment}
\usepackage{color}
\makeatletter
\renewcommand{\todo}[1]{%
  \ifinfloat{\pdfcomment[hoffset=-5pt,icon=Help]{#1}}
            {{\@todo{\pdfcomment[hoffset=-5pt,icon=Help]{#1}}}}}
\renewcommand{\todo}[1]{}
\makeatother

\usepackage{siunitx}
\usepackage[version=3]{mhchem}

\renewcommand{\figurename}{Figure}
\renewcommand{\tablename}{Table}
\makeatletter
\renewcommand{\fnum@figure}{\textbf{\figurename~\thefigure}}
\renewcommand{\fnum@table}{\textbf{\tablename~\thetable}}
\makeatother

\makeatletter
\newcommand\na[3][\@empty]{%
  {%
    \ifx#1\@empty
      \lowercase{\def\short{\textsc{#2}}}
    \else
      \def\short{#1}
    \fi
    \expandnext{\newacronym{#2}}{\short}{#3}
  }
}
\makeatother
\na{QMC}{Quantum Monte Carlo}
\na{CM}{Center of Mass}
\na{TF}{Thomas-Fermi}
\na{FNQMC}{Fixed-Node \gls{QMC}}
\na{AFQMC}{Auxiliary Field \gls{QMC}}
\na{DFT}{Density Functional Theory}
\na{DMFT}{Density-Matrix Functional Theory}
\na{KS}{Kohn-Sham}
\na{LDA}{Local Density Approximation}
\na{BCS}{Bardeen-Cooper-Schrieffer}
\na[\textsc{b}d\textsc{g}]{BdG}{Bogoliubov-de Gennes}
\na{SLDA}{Superfluid Local Density Approximation}
\na{ETF}{Effective Field Theory}
\na{UFG}{Unitary Fermi Gas}
\na{UNEDF}{Universal Nuclear Energy Density Functional}
\na[\textsc{s}\textup{ci}\textsc{dac}]
   {SciDAC}{Scientific Discovery through Advanced
  Computing}
\na[\textsc{d}\textup{o}\textsc{e}]{DoE}{Department of Energy}
\na{LDRD}{}
\na{LANL}{Los Alamos National Laboratory}
\na{NERSC}{National Energy Research Science Computing}
\na{EMMI}{ExtreMe Matter Institute}
\na{DFG}{}
\na{GSI}{}
\newcommand{\MMFGRANT}{\textsc{de-fg02-00er41132}}
\newcommand{\SGGRANTa}{\textsc{de-fc02-07er41457}}
\newcommand{\SGGRANTb}{\textsc{de-ac52-06na25396}}
\newcommand{\AGGRANTa}{\textsc{de-fg02-97er41014}}
\newcommand{\AGGRANTb}{\textsc{de-ac52-06na25396}}

\newcommand\exclude[1]{}

\renewcommand{\d}{\mathrm{d}}
\providecommand{\abs}[1]{\lvert{#1}\rvert}
\providecommand{\norm}[1]{\lVert#1\rVert}
\newcommand{\vect}[1]{\vec{#1}}
\DeclareRobustCommand{\order}{\ensuremath{\mathcal{O}}}
\DeclareMathOperator{\sech}{sech}
\makeatletter
\newcommand{\ifinfloat}[2]{\ifnum\@floatpenalty<0\relax #1\else #2\fi}
\makeatother

\newcommand\intd[2]{\int\d^{#1}{#2}\;}
\newcommand\intdbar[2]{\int\frac{\d^{#1}{#2}}{(2\pi)^{#1}}\;}

\newcommand\braket[1]{\langle #1\rangle}
\newcommand\op[1]{\hat{#1}}
\newcommand\mat[1]{\bm{#1}}
\newcommand\mR{\mat{\rho}}

\newcommand\mH{\mat{H}}

\newcommand\sumint[2]{
  \sum\hspace{-1.2em}\int\;\frac{\d^{#1}{#2}}{(2\pi)^{#1}}\;}

\usepackage[utf8]{inputenc}

\begin{document}

\title{Effective-Range Dependence of Resonantly Interacting Fermions}

\author{
  Michael McNeil Forbes,$^{1,2}$
  Stefano Gandolfi,$^{3}$ 
  and Alexandros Gezerlis$^{2,4,5}$}

\affiliation{$^1$Institute for Nuclear Theory, University of Washington,
  Seattle, Washington 98195--1550 USA}
\affiliation{$^2$Department of Physics, University of Washington, Seattle,
  Washington 98195--1560 USA}
\affiliation{$^3$Theoretical Division, Los Alamos National Laboratory, Los
  Alamos, New Mexico 87545, USA}
\affiliation{$^4$\gls{EMMI}, \acrshort{GSI} Helmholtzzentrum für
  Schwerionenforschung GmbH, 64291 Darmstadt, Germany}
\affiliation{$^5$Institut für Kernphysik, Technische Universität
  Darmstadt, 64289 Darmstadt, Germany}

\date{\today}

\begin{abstract}
  \noindent
  We extract the leading effective range corrections to the equation of state of
  the unitary Fermi gas from \textit{ab initio} \gls{FNQMC} calculations in a
  periodic box using a \gls{DFT}, and show them to be universal by considering
  several two-body interactions.  Furthermore, we find that the \gls{DFT} is
  consistent with the best available unbiased \gls{QMC} calculations, analytic
  results, and experimental measurements of the equation of state.  We also
  discuss the asymptotic effective-range corrections for trapped systems and
  present the first \gls{QMC} results with the correct asymptotic scaling.
\end{abstract}
\preprint{\textsc{la-ur-12-21036}}
\preprint{\textsc{int-pub-12-022}}
\pacs{
  67.85.-d,   
  71.15.Mb,   
  31.15.E-,   
  03.75.Ss,   
  24.10.Cn,   
  03.75.Hh,   
  21.60.-n    
}

\maketitle
\glsresetall
\lettrine{T}{he fermion many-body problem} plays a fundamental role in a vast
array of physical systems, from dilute gases of cold atoms to nuclear physics in
nuclei and neutron stars.  The universal character of this problem -- each
system is governed by a similar microscopic theory -- coupled with direct
experimental access in cold atoms, has led to an explosion of recent interest
(see Refs.~\cite{IKS:2008, *giorgini-2007, *Zwerger:2011} for reviews).  Despite
this broad applicability, we still do not fully understand even the simplest
system: the ``unitary gas'' comprising equal numbers of two fermionic species
with a resonant $s$-wave interaction of infinite scattering length
$a_{s} \to \infty$.  Lacking any scale beyond the total density $n_{+} = n_a +
n_b$, the unitary gas admits no perturbative expansion and requires experimental
measurement or accurate numerical simulation for a quantitative description.
Typical \gls{QMC} calculations, however, can access at most a few hundred
particles, while experiments can measure only a handful of properties.
\Gls{DFT} provides a complementary approach through which one may extrapolate
these results to large systems beyond the reach of direct simulation.  The
question of how the unitary gas approaches the thermodynamic limit has also been
studied in~\cite{Morris:2010, FGG:2010, Li:2011, Braun:2011}.

In this paper, we consider the effects of a finite effective range $r_e$ on the
unitary gas.  Our motivation is two-fold. First, neutron matter -- well
approximated by a unitary gas~\cite{Gezerlis;Carlson:2008-03} -- differs
primarily due to a finite range.  Characterizing the finite range effects
therefore have physical relevance.  Second, we wish to directly use a \gls{DFT}
-- a finite-range version of the \gls{SLDA} -- to fit \gls{QMC} simulations and
extract thermodynamic properties without having to first extrapolate to zero
range as was done in~\cite{FGG:2010}.  Directly fitting the finite range
\gls{QMC} data provides a much more stringent test of the \gls{SLDA}.  We use
this finite-range \gls{DFT} to extrapolate to the thermodynamic limit the linear
range dependence of the equation of state, and demonstrate its universality by
simulating three different potentials.  Two of the potentials include a
repulsive core to address issues of contamination by deep bound states.  We also
show that the \gls{SLDA} consistently fits all available unbiased zero-range
\textit{ab initio} results for the symmetric unitary gas. Finally, we present
\gls{QMC} results for trapped systems that demonstrate the correct asymptotic
scaling as predicted by the low energy effective theory for the unitary gas.

Here we consider symmetric $T=0$ systems comprising equal numbers of two neutral
Fermi species with equal mass with a short-range interaction.
These systems are directly realized by two of the lowest lying hyperfine states
of \ce{^6Li} in cold atomic systems, and approximately realized in dilute
neutron-rich matter in the crusts of neutron stars. At sufficient dilution, the
interaction can be characterized by the two-body s-wave phase shifts through the
effective range expansion (see for example~\cite{Bethe:1949})
\begin{gather}
  \label{eq:kcotdelta}
  k \cot\delta_k = \frac{-1}{a_s} + \frac{r_ek^2}{2} + \order(k^4)
\end{gather}
where $a_s$ is the s-wave scattering length and $r_e$ is the effective range.
The zero-range unitary limit is realized when the scattering length is tuned
$a_s \rightarrow \infty$ and the system is diluted such that $kr_e \rightarrow
0$: this is referred to as the symmetric \gls{UFG}.

The lack of scales implies that the symmetric \gls{UFG} is
fully characterized by the universal Bertsch parameter~\cite{mbx,
  *Baker:1999:PhysRevC.60.054311, *baker00:_mbx_chall_compet} $\xi_S =
\mathcal{E}/\mathcal{E}_{FG}$, where $\mathcal{E}_{FG} = (3/5) n_{+} E_{F}$ is the
energy density of a free Fermi gas with the same total density $n_{+} =
k_F^3/(3\pi^2)$, and $E_{F} = \hbar^2 k_F^2/2m$ is the Fermi energy.

In \ce{^6Li} cold-atom experiments (see~\cite{Ku:2011} for details), $a_s
\approx \infty$ can be tuned using the wide magnetic Feshbach resonance at
$\SI{834.1+-1.5}{G}$~\cite{Bartenstein:2005uq} with an effective range of
$r_e\approx \SI{4.7}{nm}$, while the gas can be cooled at densities of $1/k_F
\approx \SI{400}{nm}$ so that $k_Fr_e \approx 0.01$.  In dilute neutron matter
$a_{nn}\approx\SI{-18.9(4)}{fm}$~\cite{Gonzalez-Trotter:2006, *Chen:2008} and
$r_{nn}\approx \SI{2.75(11)}{fm}$~\cite{Miller:1990}, while densities are on the
order of $1/k_F \sim \SI{1}{fm}$: thus, $k_Fr_e \approx 3$ is several orders of
magnitude larger than in cold-atom systems.

Although there are formal ways of dealing with the divergences introduced by the
zero-range limit (see~\cite{Tan:2005uq,*Tan:2008kx,*Tan:2008uq} for an
interesting approach), most \textit{ab initio} calculational techniques require
an explicit regulator in the form of a finite-range potential or a lattice
cutoff. To extract the unitary parameters thus requires an extrapolation to zero
effective range.  Range effects in the \gls{UFG} are also discussed
in~\cite{Schwenk:2005ka} (large $r_e$), \cite{Li:2011} (\gls{FNQMC}), and
in~\cite{Simonucci:2011, Caballero-Benitez:2012} (\gls{BdG} approximation).

\section{Summary}
\noindent
Here we present a summary of our results.  We use a variational \gls{FNQMC}
algorithm to find upper bounds on the energy for systems of $4$ to $66$
particles in a periodic box for a variety of effective ranges $k_Fr_e
\lessapprox 0.3$ and for different potentials with the same scattering length
$a_s = \infty$ and range.  We fit these directly with a modified \gls{DFT} that
models the range dependence in order to extrapolate to the thermodynamic an
upper bound on the Bertsch parameter $\xi$, and the leading order universal
effective range dependence $\zeta_e$:
\begin{subequations}
  \begin{gather}
    \xi(k_Fr_e) \lessapprox \xi^* + \zeta_e^* k_F r_e + \order(k_Fr_e)^2
    \nonumber\\
    \begin{aligned}
      \xi^* &= 0.3897(4) & \zeta_e^* &= 0.127(4).
    \end{aligned}
  \end{gather}
  By comparing several potentials, we confirm that these are indeed universal.
  The \gls{FNQMC} results contain a systematic error due to the variational
  nature of the method.  To better understand this, we also fit with the
  \gls{DFT} a collection of unbiased exact, \gls{QMC}, and experimental results
  for systems with $2$ to $10^6$ particles, obtaining a best fit of
  \begin{align}
    \xi_S &= \num{0.3742 +- 0.0005}.
  \end{align}
  We also demonstrate for the first time, \gls{QMC} results for trapped systems
  that exhibit the correct asymptotic behavior in the thermodynamic limit. 
\end{subequations}

\section{\ACRshort{QMC} Model}
\noindent
We use a \gls{FNQMC} algorithm to simulate the Hamiltonian
\begin{gather}\label{eq:H}
  \mathcal{H} = \frac{\hbar^2}{2m}\Biggl(
  - \sum\limits_{k = 1}^{N_+} \nabla_k^{2}
  \;-\; \sum_{i,j'}V(r_{ij'})\Biggr),
\end{gather}
where $V(r)$ is an inter-species interaction (off-resonance intra-species
interactions are neglected).  The \gls{FNQMC} algorithm projects out the state
of lowest energy from the space of all wave functions with fixed nodal structure
as defined by an initial many-body wave function (ansatz).  By varying the
Ansatz, we obtain an upper bound on the ground-state energy.

We use the trial function introduced in~\cite{CCPS:2003}:
\begin{gather*}
  \Psi_T = 
  \mathcal{A}[
  \phi(\mathbf{r}_{11'})\phi(\mathbf{r}_{22'})\cdots\phi(\mathbf{r}_{nn'})]
  \prod_{ij'}f(r_{ij'}),
\end{gather*}
where $\mathcal{A}$ antisymmetrizes over particles of the same spin (either
primed or unprimed) and $f(r)$ is a nodeless Jastrow function introduced to
reduce the statistical error. The antisymmetrized product of $s$-wave pairing
functions $\phi(\mathbf{r}_{ij'})$ defines the nodal structure:
\begin{gather*}
  \phi(\mathbf{r}) = 
  \sum_{\mathbf{n}}\alpha_{\norm{\mathbf{n}}}e^{ i {\mathbf{k}}_{\mathbf{n}} \cdot
    \mathbf{r}}
  +
  \tilde{\beta} (r).
\end{gather*}
The sum is truncated (we include ten coefficients) and the omitted short-range
tail is modelled by the phenomenological function $\tilde{\beta}(r)$ chosen to
ensure smooth behavior near zero separation.  We use the same form for
$\tilde{\beta}(r)$ as in~\cite{Gandolfi:2010a} and vary the 10 coefficients
$\alpha_{\norm{\mathbf{n}}}$ for each $N_+$ and for each different two-body
potential to minimize the energy as described in Ref.~\cite{Sorella:2001}.  The
same ansatz suffices for different effective ranges, but an independent
optimization is required for each $N_+$.

\begin{figure}[bp]
  \centering
  \includegraphics[width=\columnwidth]{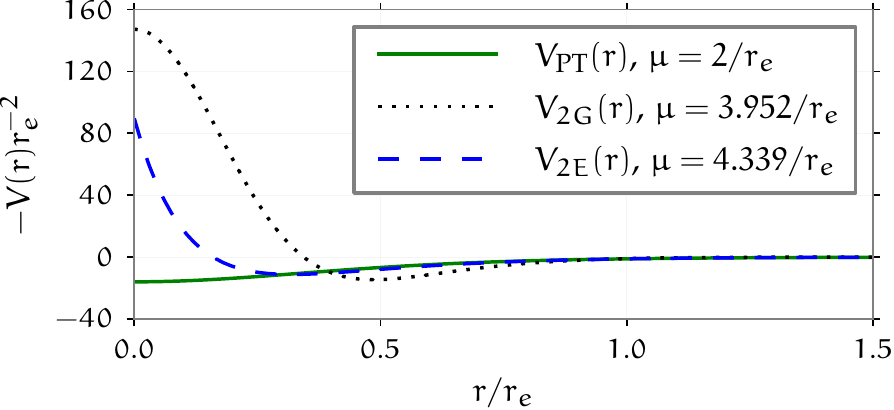}
  \caption{(color online) Finite range potentials~(\ref{eq:Vs}) used in the
    Hamiltonian~(\ref{eq:H}) for our \gls{QMC} bounds.}
  \label{fig:V}
\end{figure}

We compare the following potentials:
\sisetup{round-mode=figures, round-precision=4}
\begin{subequations}
  \label{eq:Vs}
  \begin{align}
    V_{PT}(r) &= 4\mu^2\sech^2(\mu r), \label{eq:cosh}\\
    V_{2G}(r) &= \num{3.14439042914988}
    \mu^2\left(e^{-\mu^2 r^2/4} - 4e^{-\mu^2 r^2}\right),\label{eq:2Gauss}\\
    V_{2E}(r) &= \num{4.76366047186652}
    \mu^2\left(e^{-\mu r} - 2e^{-2\mu r}\right).\label{eq:Morse}
  \end{align}
\end{subequations}
These potentials are all tuned to have infinite two-body $s$-wave scattering
length.  The first potential~(\ref{eq:cosh}) is of the
modified--P\"{o}schl-Teller type; the second~\eqref{eq:2Gauss} and
third~\eqref{eq:Morse} potentials have a repulsive core.  When tuned to
unitarity, the effective range $r_e$ is proportional to $\mu^{-1}$ as shown in
figure~\ref{fig:V}.

One criticism of purely attractive potentials -- including the widely used
modified--Pöschl-Teller potential~\eqref{eq:cosh} -- is that they may contain
deeply bound states where many particles lie within the range of the potential.
Formally, the ground state is thus not the universal dilute \gls{UFG}, but some
tightly bound state that is highly sensitive to the range.  In principle, this
state may contaminate the variational \gls{QMC} calculation, but in practice,
there is insufficient overlap between the variational wave function and this
deep bound state.  (Simulations longer by several orders of magnitude would be
required to see the influence of such low-energy states.)

\begin{figure}[tbp]
  \centering
  \includegraphics[width=\columnwidth]{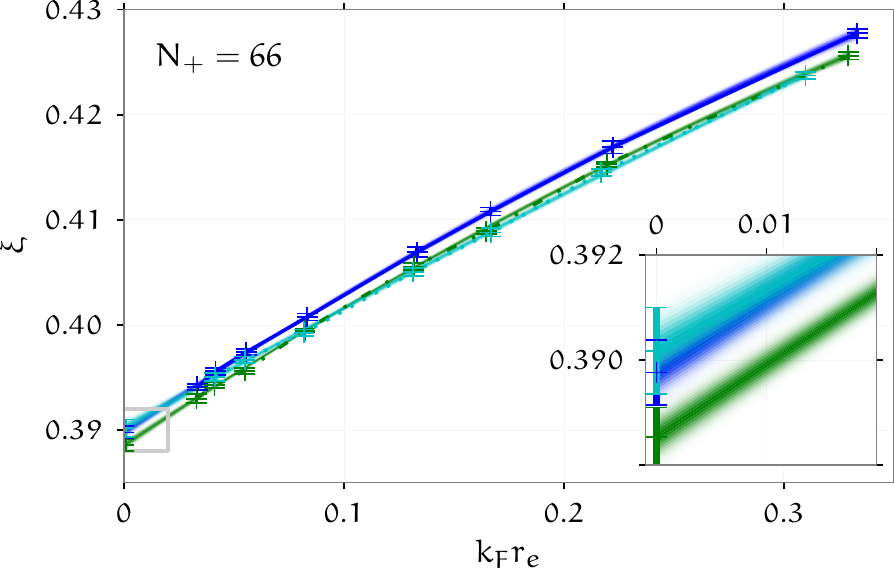}
  \caption{(color online) Effective range dependence of the ground-state
    energy-density $\xi(k_Fr_e) = \mathcal{E}/\mathcal{E}_{FG}$ of $N_+=66$
    fermions in a periodic cubic box in the unitary limit.  The points with
    error-bars are the raw \gls{QMC} results and the bands are the $1\sigma$
    error bands of polynomial fits. The upper (blue) curve that extrapolates to
    $\xi = \num{0.3898(5)}$ is the new quadratic fit to the
    modified--P\"{o}schl-Teller potential~\eqref{eq:cosh}.  The middle (green)
    curve that extrapolates to $\xi = \num{0.3885(5)}$ is the quadratic fit for
    the new double-Gaussian potential~\eqref{eq:2Gauss}.  Finally, the lower
    (cyan) curve that extrapolates to $\xi = \num{0.3902(7)}$ is the quadratic
    fit to the double-exponential potential~(\ref{eq:Morse}).}
  \label{fig:comparison}
\end{figure}

The repulsive cores of \eqref{eq:2Gauss} and \eqref{eq:Morse} help allay these
concerns by reducing the possibility of contamination from deeply bound states.
We find agreement between the purely attractive P\"{o}schl-Teller potential and
these repulsive potentials, demonstrating that all three potentials may be used
to calculate properties of the \gls{UFG}, and verifying the model-independence
of the universal parameters.  We show the upper bounds for the energy of
$N_+=66$ particles at various effective ranges in figure~\ref{fig:comparison}.
For ranges less than $k_Fr_e \lesssim 0.3$ a three-parameter quadratic model is
sufficient to extrapolate to zero range without a systematic bias.  This fit is
shown in table~\ref{tab:N66-extrap} for the three potentials, and the magnitude
of the quadratic parameter can be used to estimate the linear regime.

\begin{table}[bp]
  \sisetup{round-precision=2,round-mode=places}
  \begin{tabular}{rSSSS}
    \toprule
    & {$\xi_{66}$ ($=a_0$)} & {$\zeta_e(66)$ ($=a_1$)} & {$a_2$} &
    {$\chi^2_{\text{r}}$}\\
    \midrule
    $V_{PT}$ & 0.3898+-0.0004 & 0.14+-0.01 & -0.07 & 0.2 \\
    $V_{2G}$ & 0.3885+-0.0004 & 0.14 +-0.01  & -0.08 & 0.4 \\
    $V_{2E}$ & 0.3902+-0.0005 & 0.12+-0.01 & -0.03 & 0.3 \\
    \bottomrule
  \end{tabular}
  \caption{\label{tab:N66-extrap}%
    Comparison of the zero-range extrapolations of
    $\mathcal{E}(k_Fr_e)/\mathcal{E}_{FG} = \xi_{66} + \zeta_e(66)k_Fr_e +
    a_2(k_Fr_e)^2 + \order(r_e^3)$ for 
    quadratic fits of the $N_+=66$ \gls{QMC} results.  These values
    are higher than, but consistent with the value $\zeta_e(66) = \num{0.11(3)}$
    reported in~\cite{Carlson:2011}.  The extrapolations of these parameters to
    the thermodynamic limit $N_+=\infty$ are listed as $\xi = a_0$ and $\zeta_e
    = a_1$ in the $\xi$ block of table~\ref{tab:thermo}.  We include the
    quadratic 
    coefficient simply to show that the \gls{QMC} can be fit using a linear form
    for $k_Fr_e < \epsilon_{\text{abs}} \abs{a_1/a_2}$ to an absolute
    accuracy of about $\epsilon_{\text{abs}}$: we do not have any a priori
    reason to believe that this parameter is universal.  The systematic error
    due to neglecting the cubic terms is on the same order as the quoted
    $1\sigma$ statistical errors.
  }
\end{table}

In~\cite{FGG:2010}, each $N_+$ was independently extrapolated to zero effective
range, then the unitary \gls{SLDA} \gls{DFT} was fit to the extrapolated
results.  It was claimed that a cubic fit was required to extrapolate the
results for $k_Fr_e < 0.35$ to zero range, however, the smallest ranges $k_Fr_e
< 0.1$ had a small systematic bias in the energy due to the Trotter
decomposition of the many-body propagator $e^{-\op{H}\delta\tau} \approx
e^{-\op{V}\delta\tau/2} e^{-\op{K}\delta\tau} e^{-\op{V}\delta\tau/2} +
\order(\op{V}\delta\tau)^3$ where $\delta\tau/\hbar$ is the imaginary time-step.
Since the potentials~\eqref{eq:Vs} scale roughly as $\op{V} \propto \mu^2
\propto r_e^{-2}$, for small ranges, one needs a very small imaginary time-step,
which is computationally expensive. The extrapolated values of $\xi$ were only
underestimated for the larger systems (by $\sim 3\%$), but extracting the slope
of $\xi(k_Fr_e)$ requires higher accuracy.  Here we have carefully simulated
with smaller time-steps (for the ranges considered here, $\delta\tau E_F \approx
5\times 10^{-6}$ is sufficient to avoid any bias) to find that, for $k_Fr_e
\lessapprox 0.3$, a quadratic (but not linear) fit is sufficient.  We also no
longer use an independent zero-range extrapolation for each $N_+$.  Instead, we
use a generalized finite-range--\gls{SLDA} to fit all of the
finite-range--\gls{QMC} results with a common set of parameters.  This requires
simultaneous consistency over all ranges \emph{and} all particle numbers,
providing a more rigorous test than independently extrapolating each $N_+$.

\section{\ACRshort{SLDA} \ACRshort{DFT} with Finite Range}
As was shown in~\cite{FGG:2010}, the finite-size (``shell'') effects in
$\xi_S(N_+)$ can be well modelled by a simple local \gls{DFT} for the unitary
Fermi gas, but are not even qualitatively reproduced by adding only gradient or
kinetic corrections~\cite{Papenbrock:2005fk, Rupak:2008fk, Salasnich:2008}.  In
this paper we retain the same three-parameter form originally introduced in
Ref.~\cite{Bulgac:2007a} (called the \gls{SLDA}), but present a simple
generalization that accounts for finite-range effects.  With this generalized
form, we can directly fit the \gls{QMC} results without the need to first
extrapolate to zero range.  We first briefly review the form of the \gls{SLDA}
\gls{DFT}, then discuss the finite-range generalization.

The \gls{SLDA} \gls{DFT} is formulated in terms of three local densities
(see~\cite{Bulgac:2011} for a review): the total density $n_+$, the total
kinetic density $\tau_+$, and an anomalous $\nu$:
\begin{align*}
  n_+ &=
  2\sum_{n}\abs{v_{n}}^2 \sim \braket{\op{a}^\dagger\op{a}} +
  \braket{\op{b}^\dagger\op{b}},
  \\
  \tau_+ &= 2\sum_n\abs{\nabla v_n}^2 \sim
  \braket{\vect{\nabla}\op{a}^\dagger\cdot\vect{\nabla}\op{a}} +
  \braket{\vect{\nabla}\op{b}^\dagger\cdot\vect{\nabla}\op{b}},
  \\
  \nu &= \sum_{n} u_{n}v_{n}^{*}\sim \braket{\op{a}\op{b}}.
\end{align*}
These are expressed in terms of the Bogoliubov quasiparticle wave functions
$u_{n}(\mathbf{r})$ and $v_{n}(\mathbf{r})$ -- sometimes called ``coherence
factors''.

The three-parameter \gls{SLDA} may then be expressed as
\begin{gather*}
    \mathcal{E}_{\gls{SLDA}} =
    \frac{\hbar^2}{m}\left(
      \frac{\alpha}{2}\tau_{+} +
      \beta \frac{3}{10}(3\pi^2)^{2/3}n_{+}^{5/3}\right) +
    g\nu^{\dagger}\nu,
\end{gather*}
where $\alpha = m/m_{\text{eff}}$ parametrizes the inverse effective mass;
$\beta$ parametrizes the self-energy; and $g$ parametrizes the pairing
interaction.  In the presence of pairing, the local kinetic and anomalous
densities are divergent
\begin{subequations}
  \begin{align*}
    \lim_{\delta\rightarrow 0}\nu(\vect{x}, \vect{x}+\vect{\delta}) 
    &\rightarrow \frac{A_\nu}{\delta} + \nu_r(\vect{x}) + \order(\delta),\\
    \lim_{\delta\rightarrow 0}\tau_+(\vect{x}, \vect{x}+\vect{\delta}) 
    &\rightarrow \frac{A_\tau}{\delta} + \tau_r(\vect{x}) + \order(\delta),
  \end{align*}
\end{subequations}
where $A_\nu$, $A_\tau$, $\nu_r$ and $\tau_r$ are finite. One must regulate the
theory if one wishes to maintain a local formulation, which greatly simplifies
the computational aspects of the \gls{DFT}.  The most general form of a local
functional involving these three densities is a function of these four finite
quantities, but restricting the form to bounded functionals is somewhat
non-trivial~\cite{Forbes:2012a}, and we shall not consider these generalizations
here.

We note that the $1/\delta$ divergence corresponds to a long $1/k^2$ momentum
tail in the Fourier transform of the anomalous and kinetic densities.  This
follows from the short-range nature of the potential as has been emphasized by
Tan~\cite{Tan:2005uq,*Tan:2008kx,*Tan:2008uq}.  The most straightforward route
is to simply introduce a momentum cutoff $k<k_c$ and then define the theory in
the limit of large cutoff.  The local densities then behave as
\begin{align*}
  \tau_+ &= A_\tau\Lambda + \tau_r
  + \order(\Lambda^{-1}), 
  &
  \nu &= A_\nu\Lambda + \nu_r
  + \order(\Lambda^{-1}),
\end{align*}
where $\Lambda = \int k^{-2}\d^3{k}/(2\pi)^3 = k_c/2\pi^2$.  Within the
single-particle framework of the \gls{DFT}, these are related to the gap
$\Delta$: $A_\tau = 2m\abs{\Delta}^2/\alpha^2$, and $A_\nu = \Delta/\alpha$.
Similar short-range behavior is expected in the physical density distributions
where the coefficients $A_\tau$ and $A_\nu$ are related to the Tan's ``contact''
$C$ -- for example, $A_\nu = \sqrt{2C}/2m$ -- and it is tempting to interpret
$\sqrt{2C} \approx 2m\Delta/\alpha$ as a prediction of the \gls{DFT}, especially
at unitarity where they seem to be related numerically.  This cannot hold in
general: in particular, the contact $C$ is related to the short-range nature
of the interaction and persists in the normal phase (either meta-stable or above
the critical temperature $T>T_c$) where the order parameter $\Delta$
vanishes~\cite{Tan:pc}. The inverse coupling constant may be expressed
\begin{gather*}
  g^{-1} = n_{+}^{1/3}/\gamma - \Lambda/\alpha,
\end{gather*}
where $\gamma$ is the third dimensionless parameter characterizing the
\gls{SLDA}.

The equations of motion follow by minimizing the total energy $E =
\int\d^{3}{x}\; \mathcal{E}_{SLDA}$ with respect to the occupation factors $u$
and $v$ subject to the constraints of fixed total particle number $N_+$ and
normalization.  This leads to the following single-particle Hamiltonian for the
Bogoliubov quasiparticle wavefunctions:
\begin{align*}
  \begin{pmatrix}
    K & \Delta^\dagger
    \\
    \Delta & - K
  \end{pmatrix}
  \begin{pmatrix}
    u_n\\
    v_n
  \end{pmatrix}
  &= E_n
  \begin{pmatrix}
    u_n\\
    v_n
  \end{pmatrix},
  &
  K &= \hbar^2\frac{-\vect{\nabla}\alpha\vect{\nabla}}{2m} - \mu + U
\end{align*}
where $U = \partial\mathcal{E}/\partial{n_+}$, and $\Delta = -g\nu$.  These must
be solved self-consistently to find the stationary configurations.  With
infinite cutoff, the self-consistency equations become
\begin{align*}
  U &= \beta E_F - \frac{\abs{\Delta}^2}{3n_+^{3/2}\gamma},
  &
  \Delta &= - \gamma\frac{\nu_r}{n_+^{1/3}}.
\end{align*}
The mean-field \gls{BdG} equations may be recovered by setting $\alpha=1$,
$\beta =0$, and replacing $\smash{n_+^{1/3}}/\gamma = 4\pi/a$.  The resulting functional
contains no explicit density dependence, and so contains no self-energy $U=0$.
The \gls{SLDA} differs from the \gls{BdG} equations by the inclusion of an
effective mass and a self-energy.

The \gls{SLDA} functional is defined by the three dimensionless constants
$\alpha$, $\beta$, and $\gamma$.  In practice, we use the homogeneous solution
to the gap equation in the thermodynamic limit to replace $\beta$ and $\gamma$
with the more physically relevant parameters
\begin{align*}
  \alpha &\equiv \frac{m}{m_{\text{eff}}}, &
  \xi &\equiv \frac{\mathcal{E}}{\mathcal{E}_{FG}}, &
  \eta & \equiv \frac{\Delta}{E_F}
\end{align*}
as discussed in detail in appendix~\ref{app:SLDA} [see Eq.~\ref{eq:TF_thermo}].

To extend the functional to finite range, we simply let the three parameters
$\alpha$, $\xi$, and $\eta$ depend on the dimensionless combination $k_Fr_e$.
This introduces an additional explicit density dependence in the functional
through $k_F \propto \smash{n_+^{1/3}}$ and the self-energy must be modified
accordingly.  The use of the nonlinear relationships~\eqref{eq:TF_thermo}
between the polynomial form for $\alpha(k_Fr_e)$, $\eta(k_Fr_e)$, and
$\xi(k_Fr_e)$ and the parameters of the function makes this complicated to
write down, but numerically it is straightforward to propagate these derivatives
using, for example, automatic differentiation tools such as
\textsc{theano}~\cite{Bergstra:2010}.

For the small ranges considered in this paper, we find that a quadratic
parametrization suffices:
\begin{gather*}
  \alpha, \xi, \eta = a_0 + a_1k_Fr_e + a_2 (k_Fr_e)^2.
\end{gather*}
(Including higher order terms leads to no significant improvement in the quality
of the fits.)  This finite-range--\gls{SLDA} thus has 9 independent parameters
-- the three coefficients $a_n$ for each of the parameters $\alpha$, $\xi$, and
$\eta$.  In comparison, the procedure of independently extrapolating each $N_+$
to zero range introduces $3$ new parameters \emph{for each $N_+$} in addition to
the three \gls{SLDA} parameters, effecting a significant increase in the total
number of fitting parameters.  Note also that the new fits directly use the
\gls{QMC} results -- including their sub-percent statistical errors -- rather
than the extrapolated error bar from zero-range extrapolation: thus the new
fitting procedure places the \gls{SLDA} under a significantly more stringent
test.

We only expect this extension to model the effective-range dependence in
universal regions.  In particular, a true finite-range interaction would
naturally regulate the system, eschewing the need for an additional
cutoff in the \gls{DFT}.  For example, in the mean-field approximation, the use
of a finite-range separable potential $gv_kv_q$ with decaying form-factors gives
rise to a momentum-dependent gap $\Delta_k \propto v_k$ regulating the anomalous
density at large momenta.  Introducing such a natural regulation into the
\gls{DFT}, however, will likely require the introduction of some form of
non-locality, which significantly complicates the computational aspects of the
theory.

In principle, one could also introduce a dependence on the scattering length $a$
and temperature $T$ in a similar manner, making the coefficients functions of
$k_Fa$ and $k_BT/E_F$ respectively. Unlike the case with the effective range,
however, the dependence on these parameters must be modelled for all values
since the unitary limit corresponds to $k_Fa = \pm\infty$ and $T/E_F = 0$, while
for finite $a$ and $T$, the zero-density limit (at the edge of a trapped cloud
for example) is described by $k_Fa = 0$ and $T/E_F = \infty$; hence, any
physical system close to unitarity explores virtually all values of these
functions, requiring a careful and complete characterization.

\section{Results}
\subsection{Box}
\noindent
The results of this 9-parameter fit to the \gls{QMC} data-points with effective
ranges $0.03 < k_Fr_e \leq 0.33$ are shown in table~\ref{tab:thermo}.  The fit
to 60 points with $4 \leq N_+ \leq 130$ for the $V_{2G}$ potential has a reduced
chi squared $\chi^2_{\text{r}} = 5$.  The fit to 70 points for $4 \leq N_+ \leq
130$ to the $V_{PT}$ potential has $\chi^2_{\text{r}} = 7$.  We suspect that
this is due to approximating the effective range dependence with a purely local
functional as discussed earlier.
\begin{table}[htbp]
  \def\midspace{\addlinespace[\defaultaddspace]}
  \sisetup{table-text-alignment=center}
  \begin{tabular}{rSSSS[round-precision=1]}
    \toprule
    \multicolumn{1}{r}{Linear} 
    & {$a_{0}$} & {$a_{1}$} & & {$\chi^2_r$}\\
    \midrule
    $\xi_{PT}$ & 0.3911+-0.0004 & 0.111+-0.003 && 7.91932 \\
    $\xi_{2G}$ & 0.3900+-0.0003 & 0.111+-0.002 && 6.03991 \\
    \midspace
    $\eta_{PT}$ & 0.90+-0.01 & -0.85+-0.07\\
    $\eta_{2G}$ & 0.875+-0.008 & -0.82+-0.04\\
    \midspace
    $\alpha_{PT}$ & 1.303+-0.010 & -0.71+-0.08\\
    $\alpha_{2G}$ & 1.289+-0.007 & -0.69+-0.03\\

    \addlinespace[1.5\defaultaddspace]
    \multicolumn{1}{r}{Quadratic}
    & {$a_{0}$} & {$a_{1}$} & {$a_2$} & {$\chi^2_r$}\\
    \midrule
    $\xi_{PT}$ & 0.3903+-0.0007 & 0.121+-0.010 & 0.00+-0.03 & 6.87338 \\
    $\xi_{2G}$ & 0.3890+-0.0004 & 0.128+-0.004 & -0.06+-0.01 & 4.51771 \\
    \midspace
    $\eta_{PT}$ & 0.99+-0.03 & -2.1+-0.4 & 3+-1\\
    $\eta_{2G}$ & 0.879+-0.007 & -0.84+-0.03 & 0.00+-0.03\\
    \midspace
    $\alpha_{PT}$ & 1.34+-0.02 & -1.6+-0.4 & 5+-2\\
    $\alpha_{2G}$ & 1.292+-0.007 & -0.73+-0.06 & 0.1+-0.2\\
    \bottomrule
  \end{tabular}
  \caption{Best fit \gls{SLDA} parameters for linear (quadratic) 6-parameter
    (9-parameter) models: coefficients $a_0$, $a_1$, (and $a_2$) for each
    parameter $\xi$, $\eta$, and $\alpha$.  Note that the parameters $\alpha$
    and $\eta$ should be positive, requiring positive higher-order terms that
    are not properly constrained by our \gls{QMC} which only simulates $k_Fr_e
    \lesssim 0.3$: larger ranges require higher-order polynomials (or a
    different model).} 
  \label{tab:thermo}
\end{table}

As before~\cite{FGG:2010}, the best fit gap parameter $\eta$ and inverse
effective mass $\alpha$ are inconsistent with the values $\eta = 0.50(5)$ and
$\alpha = 1.09(2)$ obtained from the $N_+=66$ \gls{QMC} quasiparticle dispersion
relation~\cite{Carlson:2005kg, BF:2008}, and the values $\eta =
0.45(5)$~\cite{Carlson;Reddy:2008-04} and $\eta =
0.44(3)$~\cite{Schirotzek;Shin;Schunck;Ketterle:2008-08} extracted from
experimental data.  As we shall see below (see Eq.~\ref{eq:BestSLDA}), this may
be due to the fixed-node approximation.

Werner and Castin~\cite{Werner:2012, *Castin:2011, *Werner:2010} showed that the
many-body energy density depends linearly on the effective range in the
zero-range limit
\begin{gather}
  \label{eq:reff-dependence}
  \frac{\mathcal{E}}{\mathcal{E}_{FG}} = 
  \xi_S + \zeta_ek_Fr_e + \order((k_Fr_e)^2)
\end{gather}
where the coefficient $\zeta_e$ is a universal constant within Galilean
invariant continuous space models.  The value for this coefficient was first
estimated $\zeta_e = 0.046(7)$~\cite{Bhattacharyya:2006} by
fitting~\eqref{eq:reff-dependence} to the exact two-particle solution in a
trap.

The value for this coefficient for $N=66$ particles $\zeta_e(66) =\num{0.11(3)}$
was calculated using \gls{AFQMC} in~\cite{Carlson:2011} (see
table~\ref{tab:N66-extrap} for comparison) and is likely independent of other
universal parameters such as $\xi$ or the contact $C$~\cite{Tan:pc}.  The
finite-range--\gls{SLDA} allows us to extrapolate this result to the
thermodynamic limit (see table~\ref{tab:thermo}) where we find $\zeta_S =
0.127(4)$ by averaging the linear $\xi$ coefficient $a_1$ for both $V_{PT}$
($\zeta_e=\num{0.121(10)}$) and $V_{2G}$ ($\zeta_e=\num{0.128(4)}$) results.
Note that these are consistent, demonstrating the universality of this
coefficient.

\begin{figure*}[thbp]
  \centering
  \includegraphics[width=\textwidth]{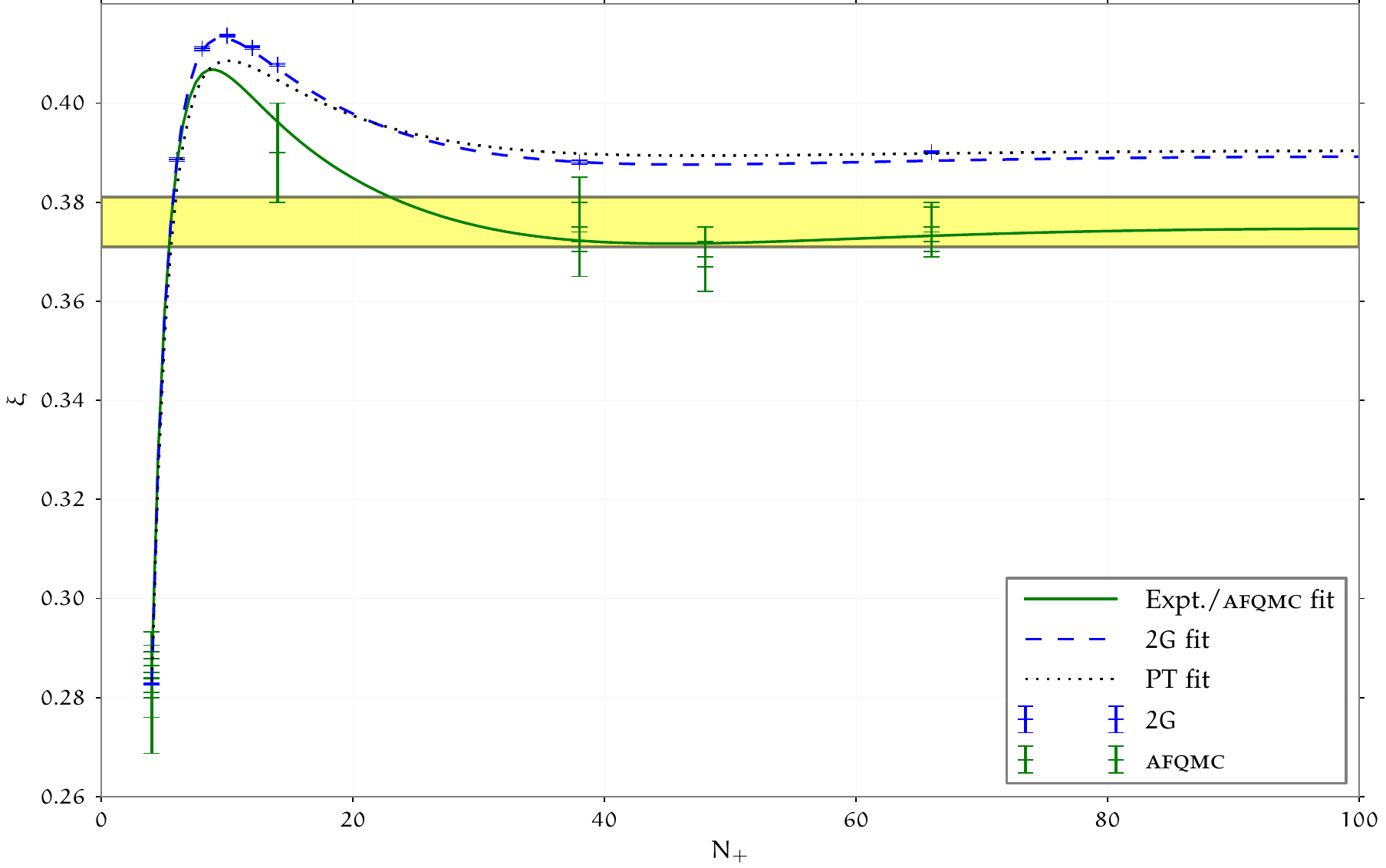}
  \caption{(color online) Comparison of \gls{SLDA} fits at zero range with
    zero-range extrapolated \gls{QMC} upper bounds (blue) with all unbiased
    zero-range extrapolations (green) from~\cite{Bour:2011,Carlson:2011} listed
    in table~\ref{tab:unbiased}.  The light (yellow) band is the experimental
    value of $\xi_{S}$~\cite{Ku:2011}.  In addition, we fit the exact
    $\xi_2=-0.4153\cdots$ value discussed in appendix~\ref{sec:particles-box}
    (not shown in the plot).  Note that this comparison allows one to assess the
    \gls{FNQMC} bound, which is tight for $N_+ \leq 6$.}
  \label{fig:all}
\end{figure*}

Unfortunately, since the \gls{FNQMC} can only provide an upper bound on the
energy, $\xi$ is systematically overestimated due to the nodal constraint.  An
improved nodal structure would lower all energies, however, and there is no a
priori reason to suspect as large a bias for $\zeta_e$.

To address the potential systematic error introduced by the fixed-node
approximation, we apply the same analysis to the recent unbiased calculations
and measurements shown in table~\ref{tab:unbiased}.  These include zero-range
extrapolations of two exact diagonalizations for $N_+=4$~\cite{Bour:2011},
zero-range extrapolations of \gls{AFQMC} results for $N_+=4$~\cite{Bour:2011}
and for $N_+\in\{4, 14, 38, 48, 66\}$~\cite{Carlson:2011}, and experimental
measurements of \ce{^6Li} for $N_+\approx 10^6$~\cite{Ku:2011}.  (Although not
strictly at zero-range, the error induced by the non-zero range in the \ce{^6Li}
experiments should be less than $0.003$ (see also section~\ref{sec:HO}).)

We use these points to fit our three-parameter zero-range \gls{SLDA}, finding:
\begin{align}
  \label{eq:BestSLDA}
  \xi_S &= \num{0.3742 +- 0.0005}, &
  \alpha &= \num{1.104 +- 0.008}, &
  \eta &= \num{0.651 +- 0.009}.
\end{align}
These error estimates must be taken with a grain of salt since not all of the
error bars quoted in table~\ref{tab:unbiased} are $1\sigma$ normal standard
deviations.  This is reflected in the rather small $\chi^2_{\text{r}} = 0.2$ of
the fit.  The results of this full fit are shown in figure~\ref{fig:all}.

This partially addresses the suspiciously large value of $\eta$ found by fitting
\gls{FNQMC} results (see table~\ref{tab:thermo}).  It appears that a large part
of the previous discrepancy is due to the fixed-node approximation which works
well for small systems, but systematically overestimates the energy of large
systems.  (The variational wavefunction has the same number of parameters for
all system sizes, we therefore expect it to better match the simpler nodal
structure of small systems than the more complicated nodal structure of larger
systems.)  The gap $\eta$ still appears to be too large, which may be a
problem when one tries to fit odd systems.

\begin{table}[hbtp]
  \begin{minipage}{\columnwidth}
    \begin{tabular}{ccl}
      \toprule
      {$N_+$} & {$\xi_{N_+}$} & Method\\ 
      \midrule
      $2$ & {$-0.415332919\cdots$} 
      & {exact (see section~\ref{sec:particles-box})}\\
      $4$ & {
        $\num{0.288 +- 0.003}$, 
        $\num{0.286 +- 0.003}$
      } & exact diagonalization~\cite{Bour:2011}\\
      $4$ & {
        $\num{0.28 +- 0.01}$
      } & \acrshort{AFQMC}~\cite{Bour:2011}\\
      $4$ & {
        $\num{0.280 +- 0.004}$
      } & \acrshort{AFQMC}~\cite{Carlson:2011}\\
      $14$ & {
        $\num{0.39(1)}$
      } & \acrshort{AFQMC}~\cite{Carlson:2011}\\
      $38$ & {
        $\num{0.370 +- 0.005}$, 
        $\num{0.372 +- 0.002}$,
        $\num{0.380 +- 0.005}$
      } & \acrshort{AFQMC}~\cite{Carlson:2011}\\
      $48$ & {
        $\num{0.372 +- 0.003}$,
        $\num{0.367 +- 0.005}$
      } &  \acrshort{AFQMC}~\cite{Carlson:2011}\\
      $66$ & {
        $\num{0.374 +- 0.005}$,
        $\num{0.372 +- 0.003}$,
        $\num{0.375 +- 0.005}$
      } &  \acrshort{AFQMC}~\cite{Carlson:2011}\\
      $10^{6}$ & {
        $\num{0.376 +- 0.005}$
      } & experiment~\cite{Ku:2011}\\
      \bottomrule
    \end{tabular}
    \caption{Unbiased zero-range box energies.  Most are extrapolated
      \gls{AFQMC} results except as noted. The $\xi_{4}$ values are consistent
      with our upper bounds $\num{0.2839 +- 0.0003}$ ($V_{PT}$), and
      $\num{0.2829 +- 0.0003}$ ($V_{2G}$).  This agreement indicates that the
      systematic error due to the fixed-node constraint is sub-percent for $N_+
      = 4$.}
    \label{tab:unbiased}
  \end{minipage}
\end{table}

The results shown in Fig.~\ref{fig:comparison} may help understand the
finite-size effects seen in neutron matter and neutron drops.  In neutron matter
at $k_F a = -10$ the difference between the energies of $N_+=20$ and $N_+=44$
particles is roughly 12\%~\cite{Gezerlis:2008:unpublished}. The range of $k_F
r_e$ values shown in Fig.~\ref{fig:comparison} is too limited too allow an
accurate extrapolation to nuclear ranges.  Even so, simple extrapolations of the
energies of $N_+=14$ and $N_+=38$ particles to $k_Fr_e = 1.45$ using linear and
quadratic forms lead to shell effects on the order of 10-20\%, which is
consistent with the (finite scattering-length) results seen in neutron
drops~\cite{Gandolfi:2010}.

\subsection{Harmonic Traps}
\label{sec:HO}

\begin{figure}[tbp]
  \centering
  \includegraphics[width=\columnwidth]{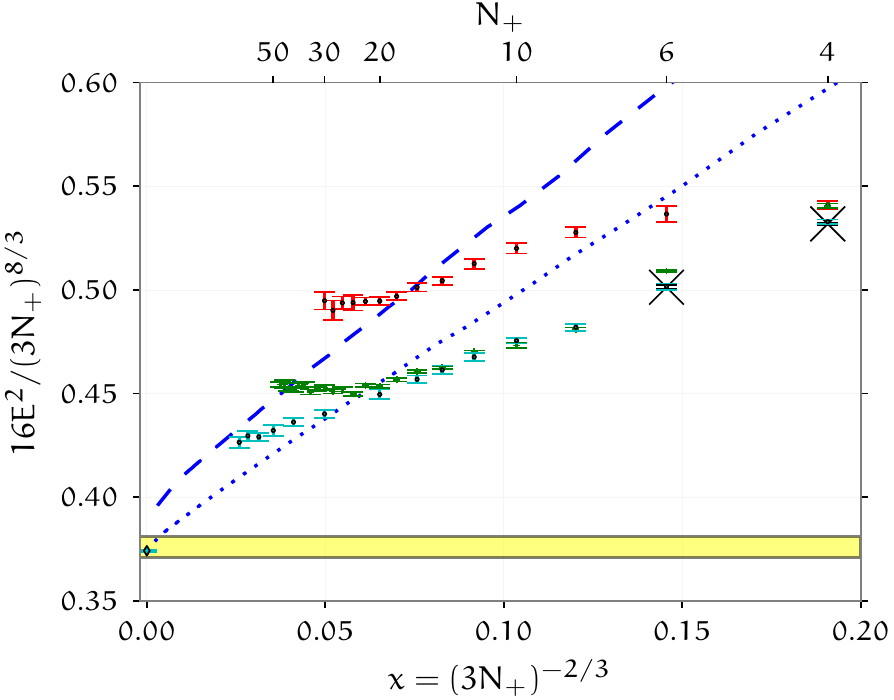}
  \caption{(color online) Ground-state energy of the harmonically trapped
    unitary Fermi gas (in units where $\hbar\omega = 1$) scaled to demonstrate
    the asymptotic form~\eqref{eq:Etrap_x} predicted by the low-energy effective
    theory of Ref.~\cite{SW:2006}.  The \gls{SLDA} with quadratic fit in
    table~\ref{tab:thermo} (dashed blue line) and unbiased
    fit~\eqref{eq:BestSLDA} (dotted blue line) is compared with zero-range
    results for $N_+\in\{4, 6\}$ from Ref.~\cite{Blume:2010} (black xs), and
    finite-range \gls{QMC} results from Ref.~\cite{Blume;Stecher;Greene:2007-12}
    (upper red dots) and Ref.~\cite{Stefano_private:2010} (middle green pluses).
    The latter have significantly lower energy, despite having a slightly larger
    effective range, suggesting that the wave functions in
    Ref.~\cite{Blume;Stecher;Greene:2007-12} were not fully optimized.  A more
    thorough optimization and extrapolation to zero effective-range yields the
    lowest points (cyan dots) which exhibit the correct scaling at large $N_+$,
    approaching the thermodynamic value of $\xi$.  We also include at $x=0$ the
    fit $\xi$ from Eq.~(\ref{eq:BestSLDA}) (cyan diamond) and the light (yellow)
    experimental band~\cite{Ku:2011}.}
  \label{fig:trap} 
\end{figure}

\noindent
As an application, we show here how the universal effective range
dependence~(\ref{eq:reff-dependence}) affects the energy of particles in an
isotropic harmonic trapping potential $V(r) = m\omega^2r^2/2$ using the \gls{TF}
approximation.  The local chemical potential is $\mu(r) = \mu_0 - V(r)$, and the
equation of state
\begin{gather*}
  \frac{\mu(r)}{E_F} = \xi_S + \frac{9}{5} \zeta_e k_F r_e + \cdots
\end{gather*} 
thereby establishes the local density and energy-density within the \gls{TF}
approximation out to the maximum \gls{TF} radius of $R =
\sqrt{2\mu_0/m}/\omega$.  Including these first two terms we thus obtain
\begin{subequations}
  \begin{align*}
    N_+ &= \frac{\omega^3 R^6}{24 \xi^{3/2}} 
    - \zeta_er_e\frac{32 \omega^4 R^7}{175\pi \xi^3} + \dots, \\
    \frac{E}{\hbar\omega} &= \frac{\omega^4 R^8}{64\xi^{3/2}} 
    - \zeta_er_e\frac{16 \omega^5 R^{9}}{225\pi\xi^3} + \dots.
  \end{align*}
\end{subequations}
In the zero-range limit, the energy of a trapped unitary gas may be calculated
using the low-energy effective theory~\cite{SW:2006} and has the form
\begin{multline*}
  E = \hbar\omega \frac{1}{4}(3N_+)^{4/3}\Bigl(
  \sqrt{\xi} + \\
  - 6\sqrt{2}\pi^2\xi(2c_1 - 9c_2)(3N_+)^{-2/3} + \order(N_+^{-7/9})
  \Bigr)
\end{multline*}
where the leading order term is the well-known \gls{TF} expression (see for
example~\cite{Chang;Bertsch:2007-08}).  The next-to-leading order term is
directly related to the $q^2$ coefficient of the static-response and the
coefficients have been estimated using the
$\epsilon$-expansion~\cite{Rupak:2008fk}.  The asymptotic corrections are due to
boundary effects beyond the validity of the effective theory.

This naturally suggests the introduction of the parameter $x=(3N_+)^{-2/3}$ so
that the asymptotic behavior of $E$ is linear in $x$.  The square of the energy
$E^2$ also exhibits linear asymptotic behavior,
\begin{gather}
  \label{eq:Etrap_x}
  \frac{16E^2}{\hbar^2\omega^2(3N_+)^{8/3}} = \xi + c x + \order(x^{7/6}),
\end{gather}
where $c= -12\sqrt{2}\pi^2\xi^{3/2}(2c_1 - 9c_2)$.  We prefer this form as $\xi$
appears as the intercept and note that the relationship is remarkably linear,
as can be seen in figure~\ref{fig:trap}.

It is interesting that, in the non-interacting system, shell-effects appear at
the same linear order $x$, leading to a fundamental uncertainty in the
coefficient $0.67 < c < 1.7$.  Pairing suppresses these shell effects, and they
are virtually non-existent in the unitary gas leading to a well-defined value of
$c$.  Note that the \gls{TF} approximation contains only the leading order term:
i.e. $c=0$.

In the \gls{TF} approximation, the leading-order effective-range correction
$\zeta_e k_Fr_e$ leads to a super-leading order (in $N_+$) correction
to~(\ref{eq:Etrap_x}):
\begin{gather}\label{eq:trap_range}
  \frac{16E^2}{\hbar^2\omega^2(3N_+)^{8/3}} = 
  \xi 
  + 
  1.17\frac{\zeta_e r_e}{\xi^{1/4}}\sqrt{\omega}x^{-1/4}
  +
  \order(r_e^2).
\end{gather}
(The coefficient is $1.17 = 2^{25/2}/1575\pi$.)  The singular $x^{-1/4}$
demonstrates that, as $N_+$ gets large, the central density becomes large and
$k_Fr_e$ corrections play an increasingly significant role.  The analysis is
therefore only valid in a limited regime where the system is sufficiently large
that the \gls{TF} approximation is valid, but where the central density is small
enough that $k_Fr_e$ remains small.  It illustrates how a finite effective range
will alter the linear asymptotic behavior expected in figure~\ref{fig:trap}.

In figure~\ref{fig:trap} we show new fixed-node \gls{QMC} results that have been
extrapolated to zero-range using a quadratic polynomial in $k_Fr_e$.  These
results represent the first \textit{ab initio} calculations to demonstrate the
correct linear asymptotic scaling as predicted by the effective theory.  In
particular, all previous results start to ``turn up'' as they approach the
thermodynamic limit.  While this is qualitatively consistent with the expected
divergent $x^{-1/4}$ behavior expected of a finite-range, the effect does not
agree quantitatively: eq.~\ref{eq:trap_range} predicts the divergence to set in
at a larger $N_+$ than seen in figure~\ref{fig:trap}.  We suspect that the
incorrect scaling of previous trapped results indicates the presence of spurious
length scales (but in principle, could also signify spurious breaking of another
symmetry).

Allowing a more flexible variational wavefunction (green
pluses~\cite{Stefano_private:2010}) improves the bound compared with the red
dots of~\cite{Blume;Stecher;Greene:2007-12}.  This seems sufficient for small
systems as witnessed by the agreement with the $N_+ \in \{4, 6\}$ results
of~\cite{Blume:2010}, but does not provide the correct asymptotic behavior in
larger traps where the density and pairing correlations differ substantially
between the center and edges of the trap.  To obtain the correct asymptotic
behavior here, we include an explicit dependence on the center-of-mass
coordinate of each pair in the variational pairing wavefunction (cyan dots).

The linear scaling of our new results indicates that this nodal approximation
does not introduce any spurious length scales, however, even with this extra
freedom, the variational bound provided for trapped systems is not as tight as
it is for homogeneous matter, and the cyan dots extrapolate to a somewhat higher
bound for the value of $\xi \approx 0.4$.  As with the homogeneous systems, we
find the same trend that the variational bound is tight for small systems, but
is less accurate for larger systems where pairing correlations become more
significant.

Finally, we have included the \gls{SLDA} predictions in the
figure~\ref{fig:trap} (blue curves) but do not use the \gls{SLDA} to fit the
results since we have not included any gradient corrections.  By construction,
the \gls{SLDA} extrapolates to the thermodynamic value of $\xi$ used in the
parametrization, but the slope is sensitive to the leading order gradient
corrections that we have neglected in this paper since they do not contribute to
homogeneous matter.  This plot contains within it hints as to the nature of the
gradient corrections to the \gls{SLDA}, but quantitative statements require
further analysis beyond the scope of this paper.

\section{Summary and Conclusions}
\noindent
In this work we have extensively analyzed the ground-state energy of strongly
interacting atoms for finite effective ranges. We present new \acrlong{FNQMC}
results for inter-atomic potentials that also contain repulsive cores: these new
potentials yield results that are statistically consistent with the purely
attractive (modified Pöschl-Teller) potential used in earlier works,
demonstrating the universality of the leading finite--effective-range
dependence, and addressing concerns about contamination of the \gls{FNQMC}
energies by deeply bound many-body states.

To model these results in a common framework, we have minimally extended the
\acrlong{SLDA} \acrlong{DFT} to directly fit the finite-range \gls{FNQMC}
results.  Although this simple generalization of the \gls{SLDA} is not
completely consistent with the \gls{FNQMC} results, it still proves to be a
useful tool for extrapolating finite-size results to the thermodynamic limit.
To assess the accuracy of the variational upper bound provided by the
\gls{FNQMC} results, we have also fit the \gls{SLDA} to unbiased
(non-variational) exact, \gls{QMC}, and experimental results from the literature
to produce a working \gls{SLDA} for modeling physical systems.  This fit
demonstrates that the three-parameter zero-range \gls{SLDA} is consistent with
the unbiased results.

Finally, we have presented new \gls{QMC} results for zero-range trapped
systems.  These results demonstrate, for the first time, the correct asymptotic
behavior in the thermodynamic limit as predicted by the low-energy effective
theory.

\glsunset{LDRD} \glsunset{UNEDF} \glsunset{SciDAC} 
\begin{acknowledgments}
  \noindent
  We thank A.~Bulgac, J.~Carlson, Y.~Castin, Y.~Nishida, and F.~Werner for
  useful discussions.  This work is supported, in part, by \textsc{us} \gls{DoE}
  grants \MMFGRANT, \AGGRANTa, \& \AGGRANTb, \gls{DoE} contracts \SGGRANTa\ 
  (\gls{UNEDF} \gls{SciDAC}) \& \SGGRANTb, by the \gls{LDRD} program at
  \gls{LANL}, by the Helmholtz Alliance Program of the Helmholtz Association
  \textsc{ha216/emmi}. Computations for this work were carried out through Open
  Supercomputing at \gls{LANL}, and at the \gls{NERSC}.
\end{acknowledgments}
\onecolumngrid
\clearpage
\twocolumngrid

\appendix
\section{Homogeneous solutions of the \acrlong{SLDA}}
\label{app:SLDA}

\noindent In this appendix, we describe some properties of homogeneous solutions
to the \gls{SLDA} functional, both in the periodic box, and in the thermodynamic
limit of infinite matter. (When these equations are applied locally at each
point in a slowly varying external potential, one obtains the \gls{TF}
approximation.) As discussed in the text, we use the thermodynamic solutions to
express the parameters $\beta$ and $\gamma$ in terms of the more physically
relevant quantities $\xi$ and $\eta$.

We start by rotating away the phase, taking $\Delta = \abs{\Delta}$ to be real.
We also note that the self-energy $U$ plays no role in the solution of the
homogeneous equations: all effects are absorbed into the effective chemical
potential $\mu_{\text{eff}}$.  One only needs to compute the self energy $U$ to
relate the effective chemical to the thermodynamic chemical potential.  Thus,
the homogeneous Hamiltonian is completely parametrized by $\alpha$,
$\mu_{\text{eff}}$, and $\Delta$.  In momentum space, the Hamiltonian is easily
diagonalized,
\begin{gather*}
  \mH = \begin{pmatrix}
    \epsilon_k & \Delta\\
    \Delta & -\epsilon_k
  \end{pmatrix}
  = \begin{pmatrix}
    u_k & v_k\\
    v_k & -u_k
  \end{pmatrix}
  \begin{pmatrix}
    E_k \\
    & -E_k
  \end{pmatrix}
  \begin{pmatrix}
    u_k & v_k\\
    v_k & -u_k
  \end{pmatrix},
  \nonumber\\
  \begin{aligned}
    \epsilon_k &= \frac{\alpha\hbar^2k^2}{2m} - \mu_{\text{eff}}, 
    &
    E_k &= \sqrt{\epsilon_k^2 + \Delta^2},
    \\
    u_k &= \sqrt{\frac{1 + \frac{\epsilon_+}{E_+}}{2}}, 
    &
    v_k &= \sqrt{\frac{1 - \frac{\epsilon_+}{E_+}}{2}}.
  \end{aligned}
\end{gather*}
In this diagonal form, the density matrix $\mR = f_\beta(\mH)$ can be computed
in a straightforward manner from the Fermi distribution function $f_\beta(\mH) =
1/[1 + \exp(-\beta \mH)]$.  For reference, the zero-temperature results are:
\begin{subequations}
  \label{eq:TF_integrals}
  \begin{align*}
    n_+(\alpha, \mu_{\text{eff}}, \Delta) 
    &= \sumint{3}{\vect{k}}\left(1 - \frac{\epsilon_k}{E_k}\right),
    \label{eq:TF_n}\\
    \frac{\tau(\alpha, \mu_{\text{eff}}, \Delta)}{2m} 
    &= \sumint{3}{\vect{k}}
    \frac{k^2}{2m}\left(1 - \frac{\epsilon_k}{E_k}\right),
    \\
    \nu(\alpha, \mu_{\text{eff}}, \Delta) 
    &= \sumint{3}{\vect{k}}\frac{\Delta}{2E_{k}}.
  \end{align*}
  The notation $\smash{\sumint{3}{\vect{k}}}$ represents either a discrete
  summation over box momenta $k_i=2\pi n_i/L_i$ or the continuous integral
  $\intd{3}{\vect{k}}/(2\pi)^3$ in the thermodynamic limit.  The regulated
  quantities $\tau_r$ and $\nu_r$ follow from these by subtracting the power-law
  divergences (this is equivalent to using dimensional
  regularization~\cite{Papenbrock:1998wb}):
  \begin{align*}
    \frac{\tau_r}{2m} 
    &= \sumint{3}{\vect{k}}
    \frac{k^2}{2m}\left(1 - \frac{\epsilon_k}{E_k}\right)
    - \intdbar{3}{\vect{k}}\frac{m\Delta^2}{\alpha^2k^2},
    \\
    \nu_r &= \sumint{3}{\vect{k}}\frac{\Delta}{2E_{k}} 
    - \intdbar{3}{\vect{k}}\frac{m\Delta}{\alpha k^2}.
  \end{align*}
\end{subequations}
Note that the subtraction integrals are continuous.  In order to implement this
regularization scheme in the periodic box, one must use a simultaneous spherical
cutoff on both the discrete and continuous momenta.  The partial sums as a
function of cutoff will fluctuate as various lattice points enter the sphere,
but the magnitude of the fluctuations will reduce and the resulting limit
converges.  Numerically, it is favorable to sum over cubic shells so that the
sequence of partial sums behaves smoothly, allowing one to accelerate the
convergence.  However, the location of the cutoff between shells must be fine
tuned to reproduce the correct result because -- unlike the spherical case --
the fluctuations never die away with a cubic cutoff.

From the integrals one can see that the effective mass can be scaled out to
define the following finite functions:
\begin{subequations}
  \label{eq:TF_functions}
  \def\args{\left(\frac{\Delta}{\alpha}, \frac{\mu_{\text{eff}}}{\alpha}\right)}
  \def\argsb{(\alpha, \mu_{\text{eff}}, \Delta)}
  \begin{align*}
    n_+\args &= n_+\argsb,
    \\
    \tilde{C}\args &= -\frac{\alpha}{\Delta}\nu_r\argsb,
    \\
    \tilde{D}\args &= \frac{\alpha^2}{\Delta^2}\tau_r\argsb.
  \end{align*}
\end{subequations}
One can thus deduce that, if the volume $V=L_xL_yL_z$ and shape of the box
$\vect{L}$ are held fixed, then the \gls{TF} equations exhibit an additional
invariance under scaling $\alpha$, $\mu_{\text{eff}}$, and $\Delta$ by the same
factor
\begin{gather*}
  \frac{\d{\alpha}}{\alpha} 
  = \frac{\d{\mu_{\text{eff}}}}{\mu_{\text{eff}}} 
  = \frac{\d{\Delta}}{\Delta}.
\end{gather*}
Note that this does not follow from dimensional analysis ($\alpha$
is already dimensionless) and expresses a non-trivial property of the \gls{TF}
equations.  These scaling relationships allow us to express everything in terms
of two dimensionless parameters -- $\aleph = \eta/\alpha$,
and the total (dimensionless) particle number $N_+ = Vn_+$ -- through the
dimensionless functions $c(\aleph, N_+)$ and $d(\aleph, N_+)$:
\begin{align*}
  &\aleph \equiv \frac{\eta}{\alpha},
  &
  &N_+ \equiv n_+ V,
  \\
  &c(\aleph, N_+) = \tilde{C}\frac{E_F^2}{\mathcal{E}_{FG}}, 
  &
  &d(\aleph, N_+) = \tilde{D}\frac{E_F^2}{\mathcal{E}_{FG}},
  \\
  &\nu_r = -\frac{3\aleph}{5} c(\aleph, N_+) n_+,
  &
  &\tau_r = \aleph^2 d(\aleph, N_+)\mathcal{E}_{FG}.
\end{align*}

In the $T=0$ thermodynamic limit $L\rightarrow \infty$ ($N_+ \rightarrow
\infty$), the integrals can be performed analytically
(see~\cite{Papenbrock:1998wb}).  We start by defining:
\begin{align*}
  k_0 &= \sqrt{\frac{2m\abs{\mu_{\text{eff}}}}{\alpha\hbar^2}}, &
  y_0 &= \frac{\mu_{\text{eff}}}{\sqrt{\Delta^2 + \mu_{\text{eff}}^2}}.
\end{align*}
We may then express our previous results as
\begin{align*}
  n_+ &=\frac{k_0^3}{3\pi^2}h_n,
  &
  \tilde{C} &= \frac{m k_0}{4\pi\hbar^2}h_c,
  &
  \tilde{D} = \frac{-\alpha^2\hbar^2k_0^{5}}{8m\pi^2\Delta^2}h_d,
\end{align*}
where the functions $h_n$, $h_c$, and $h_d$ depend only on $y_0$,
\begin{align*}
  h_n(y_0) &= \frac{3}{4}\frac{y_0f_{1/2}(y_0) - f_{3/2}(y_0)}{\abs{y_0}^{3/2}},
  &
  h_c(y_0) &= \frac{f_{1/2}(y_0)}{\pi\abs{y_0}^{1/2}},
  \\
  h_d(y_0) &= \frac{f_{5/2}(y_0) - y_0f_{3/2}(y_0)}{\abs{y_0}^{5/2}},
  &
  f_{\alpha}(y_0) &= \frac{-\pi\;P_{\alpha}(-y_0)}{\sin(\pi\alpha)},
\end{align*}
through the Legendre function $P_{\alpha}(x)$ which satisfies
\begin{align*}
  0 &= (1 - x^2)P_\alpha'' - 2xP_\alpha' + \alpha(\alpha + 1)P_\alpha,\\
  P_{\alpha}(x) &= \frac{1}{2\pi i}
  \oint \omega^{-\alpha-1} \sqrt{1-2x\omega+\omega^2}\;\d\omega.
\end{align*}
Noting that $n_+ = k_F^3/3\pi^2$ we can identify $k_F^3 = k_0^3 h_n$ and
$E_F = \hbar^2 k_F^2/2m = h_n^{2/3} k_0^2/2m =
h_n^{2/3}\abs{\mu_{\text{eff}}}/\alpha$.  We can then relate $\aleph$ directly
to $y_0$ through the monotonic function:
\begin{gather*}
  \aleph(y_0) = \frac{\Delta}{\alpha E_F}
  = \frac{\Delta}{\abs{\mu_{\text{eff}}} h_n^{2/3}(y_0)}
  = \frac{\sqrt{y_0^{-2} - 1}}{h_n^{2/3}(y_0)}.
\end{gather*}
This function has the limiting behavior:
\begin{gather*}
  \aleph = \begin{cases}
    \left(\frac{4}{1+y_0}\right)^{1/6} & \text{ where } y_0 \approx -1,\\
    \sqrt{2(1-y_0)} & \text{ where } y_0 \approx 1,
  \end{cases}
\end{gather*}
and an application of five steps of Newton's method using this as a guess
(splitting the input at the point %
\sisetup{round-precision=10}
$\aleph\approx \num{1.211292490}$ where these asymptotic forms meet) solves the inverse
problem $y_0(\aleph)$ to machine precision.  With this conversion we can
directly express
\begin{subequations}
  \begin{align*}
    c(\aleph) = c_{N_+=\infty}(\aleph) 
    &= \frac{5\pi}{8} \frac{h_c(y_0)}{h_n^{1/3}(y_0)},\\
    d(\aleph) = d_{N_+=\infty}(\aleph) 
    &= \frac{-5}{4\aleph^2}\frac{h_d(y_0)}{h_n^{5/3}(y_0)},
  \end{align*}
\end{subequations}
allowing the parameters $\alpha$, $\beta$, and $\gamma$ to be computed from the
thermodynamic values of $\alpha$, $\xi$, and $\eta$:
\begin{subequations}
  \label{eq:TF_thermo}
  \begin{align}
    \gamma &= \frac{5\alpha (3\pi^2)^{2/3}}{6c(\eta/\alpha)},
    \\
    \beta &= \xi - \frac{d(\eta/\alpha)\eta^2}{\alpha} 
    - \frac{6\eta^2\gamma}{5(3\pi^2)^{2/3}}.
  \end{align}
\end{subequations}
We use these equations to express all of our results in terms of the
thermodynamic values of $\alpha$, $\xi$, and $\eta$, even though the functional
is expressed in terms of fixed parameters $\alpha$, $\beta$, and $\gamma$.

\section{Particles in a Box}\label{sec:particles-box}
\noindent
Here we present some details about computing the energies $E$ of $N_+ = N_a +
N_b$ particles in a cubic box of size $L^3$.  There are two conventions for
expressing the energy of a box. We use $\xi_{N_+} =
\mathcal{E}(N_+)/\mathcal{E}_{FG}$ where $\mathcal{E}(N_+) = E(N_+)/L^3$. All
values of $\xi$ reported in this paper have been converted to this
normalization.  The other convention $\xi^{\text{box}} = E(N_+)/E_{FG}(N_+)$
normalizes the energy with respect to the energy of $N_+$ non-interacting
fermions in the same box (see~\cite{FGG:2010} for conversion factors).

To further constrain our fits, we include the results for $N_+=2$.  By solving
the Schrödinger equation for two particles in a periodic box of size $L^3$ with
the short-range boundary condition
\begin{gather*}
  \lim_{r \rightarrow 0}\Psi(\vect{x},\vect{x}+\vect{r})
  \propto \frac{1}{r} + k\cot\delta_k + \order(r),
\end{gather*}
one obtains 
\begin{subequations}
  \begin{align*}
    k\cot\delta_k &= 
    \frac{1}{\pi L}S\left(\left(\frac{Lk}{2\pi}\right)^2\right),\\
    S(\eta) &= \lim_{\Lambda \rightarrow \infty}
    \sum_{\vect{n}}^\Lambda\frac{1}{\norm{\vect{n}}^2 - \eta} - 4\pi \Lambda.
  \end{align*}
\end{subequations}
where $E = k^2/2m_r$ is the energy in the center-of-mass--frame and $m_r =
m/2$ is the reduced mass of the system (see for example~\cite{Beane:2003da} and
references therein).  Note that for non-interacting particles, $E_{FG}(N_+) =
0$, thus for all attractive interactions, $E \propto k^2 < 0$.  This poses no
problems since only $k^2$ enters the formulation: for example, $k\cot\delta_k$
is a series in $k^2$~(\ref{eq:kcotdelta}).

As before, the summation may be performed with partial sums over cubic shells:
these behave smoothly and are amenable to series acceleration
techniques (see~\cite{Bornemann:2004rp} for example) such as the Levin
transformation.

\begin{figure}[thbp]
  \centering
  \includegraphics[width=\columnwidth]{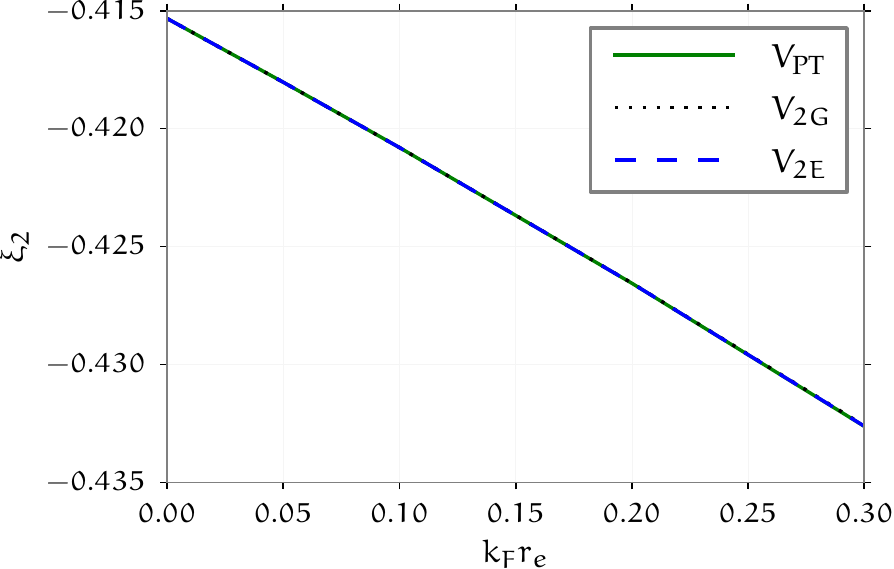}
  \caption{(color online) Exact ground-state energy $\xi_2(k_Fr_e)$ for the
    $N_+=2$ system  in a box for each of the potentials~\eqref{eq:Vs}.}
  \label{fig:two-body}
\end{figure}

The energies $\xi_2(k_Fr)$ are shown in figure~\ref{fig:two-body} for the
potentials~(\ref{eq:Vs}).  Over the ranges considered, the results are virtually
identical.  Finally, we note that the $N_+=2$ solution to the \gls{SLDA} has
only the normal solution $\Delta = 0$.  Both particles enter the $k=0$ ground
state which has zero energy, hence we can identify $\beta(k_Fr) = \xi_2(k_Fr)$.

\end{document}